\documentclass{emulateapj}

\usepackage{xspace}
\usepackage{graphicx}
\usepackage{natbib}
\usepackage{apjfonts}
\newcommand\aproxgt{\mathrel{%
     \rlap{\raise 0.511ex \hbox{$>$}}{\lower 0.511ex \hbox{$\sim$}}}}
\newcommand\aproxlt{\mathrel{%
     \rlap{\raise 0.511ex \hbox{$<$}}{\lower 0.511ex \hbox{$\sim$}}}}

\slugcomment{Submitted 2015 August 1.  Accepted $\ldots$}

\shorttitle{As above, so below: exploiting mass scaling in black hole accretion}
\shortauthors{Markoff et al.}

\begin{document}

\title{As above, so below:  exploiting mass scaling in black hole
  accretion to break degeneracies in spectral interpretation}
\author{Sera Markoff\altaffilmark{1}, Michael
  A. Nowak\altaffilmark{2}, Elena Gallo \altaffilmark{3}, Robert Hynes
  \altaffilmark{4}, J\"orn Wilms \altaffilmark{5},  Richard M. Plotkin
  \altaffilmark{3}, Dipankar Maitra \altaffilmark{6}, Catia V. 
  Silva \altaffilmark{1,7}  \& Samia Drappeau \altaffilmark{8}}
\altaffiltext{1}{Anton Pannekoek Institute for Astronomy, University
  of Amsterdam, 1098 XH Amsterdam, the Netherlands;
  S.B.Markoff@uva.nl, C.V.DeJesusSilva@uva.nl} 
\altaffiltext{2}{Massachusetts Institute of Technology, Kavli Institute for
  Astrophysics, Cambridge, MA 02139; mnowak@space.mit.edu}
\altaffiltext{3}{Dept. of Astronomy, Univ. of Michigan, Ann Arbor, MI
  48109-1042; egallo@umich.edu}
\altaffiltext{4}{Department of Physics and Astronomy, Louisiana State University, Baton Rouge, LA 70803-4001, USA; rih@redstick.phys.lsu.edu}
\altaffiltext{5}{Dr. Karl Remeis-Sternwarte \& ECAP, Universit\"at
  Erlangen-N\"urnberg, 96049 Bamberg, Germany; joern.wilms@sternwarte.uni-erlangen.de}
\altaffiltext{6}{Department of Physics and Astronomy, Wheaton College, Norton, MA 02766, USA; maitra\_dipankar@wheatoncollege.edu}
\altaffiltext{7}{SRON Netherlands Institute for Space Research, 3584 CA, Utrecht, The Netherlands }
\altaffiltext{8}{CNRS; IRAP; BP 44346, F-31028 Toulouse cedex 4,
  France; samia.drappeau@irap.omp.eu}

\begin{abstract}
  Over the last decade, the evidence is mounting that several aspects
  of black hole accretion physics proceed in a mass-invariant way.
  One of the best examples of this scaling is the empirical
  ``Fundamental Plane of Black Hole Accretion'' relation linking mass,
  radio and X-ray luminosity over eight orders of magnitude in black
  hole mass. The currently favored theoretical interpretation of this relation is
  that the physics governing power output in weakly accreting black holes
  depends more on relative accretion rate than on mass.  In order to test this theory,
  we explore whether a mass-invariant approach can simultaneously
  explain the broadband spectral energy distributions from two black
  holes at opposite ends of the mass scale but at similar Eddington
  accretion fractions.  We find that the same model, with the same
  value of several fitted physical parameters expressed in
  mass-scaling units to enforce self-similarity, can provide a good
  description of two datasets from V404 Cyg and M81*, a stellar and
  supermassive black hole, respectively.  Furthermore, only one of
  several potential emission scenarios for the X-ray band is
  successful, suggesting it is the dominant process driving the Fundamental Plane
  relation at this accretion rate.  This approach thus holds promise
  for breaking current degeneracies in the interpretation of black
  hole high-energy spectra, and for constructing better prescriptions of
  black hole accretion for use in various local and cosmological
  feedback applications.
\end{abstract}


\section{Introduction}\label{sec:intro}

\setcounter{footnote}{0} Accreting black holes, whether in Galactic
X-ray binaries (BHBs) or Active Galactic Nuclei (AGN), drive a
complicated system of inflowing (quasi-)thermalized plasma in an
accretion disk, outflowing plasma in the form of winds and/or
relativistic jets, and a hot corona that may comprise elements of both
phenomena \citep[see, e.g.][]{MarkoffNowakWilms2005}.  The basic
morphological similarities between these systems have led to the
proposal that at least some general properties of black hole (BH)
accretion might scale predictably with mass, regardless of outer
boundary conditions (i.e., fueling).

Over the last decade, there is increasing evidence for such a mapping
between BHB accretion states
\citep{McClintockRemillard2006,Belloni2010} and AGN classifications
\citep[e.g.][]{KoerdingJesterFender2006}.  The two most compelling
examples are the correspondences between variability
timescales in BHBs and AGN \citep[e.g.,][]{McHardyetal2006,
  McHardyetal2007} and the Fundamental Plane of Black Hole Activity
(hereafter FP) discovered over a decade ago
\citep{MerloniHeinzDiMatteo2003,FalckeKoerdingMarkoff2004} and
increasingly refined via several newer studies
\citep[e.g.,][]{KoerdingFalckeCorbel2006,McHardyetal2006,Gueltekinetal2009,Plotkinetal2012}.

The FP is an empirical relation between the radio and
X-ray luminosities and masses of accreting BHs in the `hard' BHB state
associated with compact, self-absorbed jets (see
\citealt{Fender2001a,McClintockRemillard2006}) and low-luminosity AGN
with jet cores: i.e., LLAGN in LINERS and FRI/BL Lacs.  Essentially all
weakly accreting AGN with jets seem to adhere to this plane.  The
planar coefficients can be derived assuming a common reservoir of
accretion power linearly dependent on accretion rate $\dot m$
(expressed in mass-scaling Eddington units
$\dot{m}=\dot{M}/\dot M_{\rm Edd}$, where
$\dot M_{\rm Edd} = L_{\rm Edd}/(0.1 c^2)$ and
$L_{\rm Edd}={4\pi G M m_p c}/{\sigma_T}$), injected into a region
whose size scales linearly with $M_{\rm BH}$, together with
conservation laws, optical depth effects and low radiative
efficiencies ($L \propto \dot m^q$, where $q\approx2$;
\citealt{FalckeBiermann1995,Markoffetal2003,HeinzSunyaev2003,Plotkinetal2012}).
The actual physics driving the FP is not yet fully understood,
primarily because of persistent degeneracy in the interpretation
of the spectral energy distributions (SEDs).  Both synchrotron
radiation as well as synchrotron self-Compton (SSC) in several flavors
of radiatively inefficient accretion flows (RIAFs;
\citealt{NarayanYi1994,YuanQuataertNarayan2003}) or outflows
\citep[e.g.,][]{MarkoffNowakWilms2005,YuanCuiNarayan2005} have
radiative efficiencies consistent with the limits set by the FP ($q\approx2$, though
see \citealt{Plotkinetal2012}).

The FP predicts that BHs regulate their power
output similarly when at similar relative accretion rates (see, e.g.,
\citealt{HeinzSunyaev2003,Markoff2010} for a broader review).  In
other words, two sources at similar $\dot m$ should radiate from
regions of similar size (in gravitational radii $r_g\equiv{GM}/{c^2}$)
and with the same physical mechanism (or at least mechanisms with the
exact same efficiencies).  This Letter explores a new approach to
quantitatively test this assumption, with an eye towards breaking the
degeneracy between synchrotron and SSC models, via the joint modeling
of broadband SEDs from two BHs at extreme ends of the mass scale.  In
Section~\ref{sec:method}, we describe the methodology and briefly
summarize the model we use for this study.  In
Section~\ref{sec:results} we present our results, and in
Section~\ref{sec:conclude} we conclude with an
outlook for potential extensions of this approach.

\section{Summary of model and methodology}\label{sec:method}

The low-energy spectrum of FP BHs consists of a
flat/inverted synchrotron component, associated with self-absorbed
emission from stratified regions along a compact jet
\citep[e.g.,][]{BlandfordKoenigl1979}.  The X-ray bands often show
evidence of weak emission from a thermal accretion disk (e.g.,
\citealt{ShakuraSunyaev1973,Mitsudaetal1984}) plus a non-thermal
component over which debate rages as to the relative contributions of
synchrotron and inverse Compton processes.  Realtime radio/X-ray
correlations in BHBs clearly demonstrate that the jets and the X-ray
source are tightly coupled over orders of magnitude in luminosity.
A mass-dependent normalization extends this relation to AGN, defining
the FP.

A straightforward test can isolate the mass-dependent effects: express
a given model in terms of mass-scaling units (i.e., all distances
expressed in $r_g$ and power in units of
$L_{\rm Edd}=1.25\times10^{38}\left(M/M_\odot\right)$ erg/s), and see
how it fares when applied to data from stellar to supermassive BHs.
This type of approach is not new: the standard thin disk paradigm
\citep{ShakuraSunyaev1973} seems to scale sensibly with mass.  The
translation of this approach to non-thermal components has not yet
been studied.  For this Letter we use the outflow-dominated model of
\cite{MarkoffNowakWilms2005} (hereafter referred to as MNW05), with
additional modifications as detailed in \cite{Maitraetal2009a}.  This
multi-scale, broadband model has been successfully applied to a
variety of BHs at both ends of the mass range individually, but
never jointly as we explore here.  We emphasize that this test
should apply for any model that can address the broadband
SEDs of weakly accreting BHs, and thus predict the FP
relations.

The details of MNW05 can be found in the above papers, and many
applications to both BHBs 
\citep[see, e.g.,][]{MarkoffNowakWilms2005,Galloetal2007,Maitraetal2009a,Plotkinetal2015},
and LLAGN in LINERS
\citep[e.g.][]{Markoffetal2001,Markoffetal2008,Maitraetal2011,Prietoetal2015}.
Here we give just basic summary of the properties of the model and the
relevant fitted parameters.

The MNW05 model includes a heuristic, multi-temperature thin disk
component \citep[e.g.,][]{ShakuraSunyaev1973,Mitsudaetal1984} whose
radius $R_{\rm in}$ and temperature $T_{\rm in}$ are fitted to the
data, and whose photons contribute to the photon field for inverse
Compton scattering.  Within $r<R_{\rm in}$ we assume that
radiatively dominant jets are anchored in a RIAF (see, e.g.,
\citealt{YuanMarkoffFalcke2002}), powered by a fraction of $\dot M c^2$ that is
divided equally between cold protons and internal pressure (radiating
leptons and magnetic fields).

A thermal particle distribution is assumed to enter the jet nozzle,
making this region something of an interface with, or proxy for, the
inner RIAF/corona. The jet flow solution is based on a
self-collimating, freely expanding hydrodynamic wind \citep[see,
e.g.,][]{FalckeBiermann1995,FalckeMarkoff2000} and thus decoupled from
the internal pressure \citep[see, e.g.,][for a relativistic
MHD-consistent treatment in development]{PolkoMeierMarkoff2014}. Thus
once conditions at the launch point are set, the scaling of physical
parameters along the jets is fully determined until the location
$z_{\rm acc}$.  There a fixed fraction of particles (60\%) is
accelerated into a power-law distribution with index $p$, and assumed
to be maintained from that point onwards by a distributed process as
implied by observations \citep[e.g.][]{Jesteretal2001}.  There is also
an option to inject particles into the jets already accelerated, in
which case $z_{\rm acc}$ is not used and a maximum Lorentz factor
$\gamma_{\rm max}$ is instead fit to the data.  The fitted parameters
are: $p$, $z_{\rm acc}$, $R_{\rm in}$ and $T_{\rm in}$, the scaled
power normalization $N_j$ (in units of $L_{\rm Edd}$) injected into
the jets at their base, of radius $r_0$ and height $h_0$ (sometimes
frozen), with ratio of magnetic to thermal gas pressure $k$ , the
temperature of the initial, mildly relativistic Maxwell-Juttner
distribution for the radiating particles $T_e$ (which also sets
$\gamma_{\rm min}$ for the injected power law case), and $f_{sc}$, a
parameter absorbing uncertainties in the acceleration efficiency when
particles are accelerated at $z_{\rm acc}$.

To compare two BHs of different masses requires SEDs of comparable,
simultaneous broadband coverage and quality.  Currently the only LLAGN
with such extensive coverage are M87 \citep{Prietoetal2015}, our
Galactic center supermassive BH Sgr A*, and M81* from a campaign
originally designed to provide a comparison source to Sgr A*. These
observations included radio (GMRT, VLA), sub-millimeter (PdBI, SMA),
and X-ray (Chandra-HETG), as described by \cite{Markoffetal2008},
where we also showed that the MNW05 model provides a good description
of the M81* SED. The fitted parameter ranges were similar to those
found in hard state BHBs; however, we were not able to break the
degeneracy between two potential origins for the X-ray emission
providing statistically comparable fits: direct synchrotron emission
from the inner jets or SSC from the jet base/corona.

To study the potential ``self-similarity'' in mass, and attempt to
break the above degeneracy, we here seek to compare the SED from M81* to 
the BHB V404 Cygni (hereafter V404), with masses $7\times10^7~M_\odot$
\citep{Devereuxetal2003} and $12~M_\odot$ \citep{Shahbazetal1994},
respectively.  We use the compiled SED of V404 from
\cite{Hynesetal2009}, where the X-ray (Chandra-ACIS), UV (Hubble Space
Telescope; HST) and radio data (VLA) were simultaneous, while optical/infrared constraints
(e.g., from Spitzer and ground based instruments) were archival.
Similarly for M81* we include archival HST (IR/UV) and Spitzer
data,  as well as ground-based constraints from ISO and MIRLIN (see
\citealt{Markoffetal2008} for details). We
apply for the first time a multi-zone, multiwavelength model jointly
to the datasets from two sources, separated by a huge dynamic range in
mass, tying together several model parameters across this mass range.
We have developed this new approach within the data analysis software
package \texttt{ISIS} \citep{Houck2002}.  Note that scale-free
parameters correspond to different physical values, therefore features
in the model SEDs corresponding to, e.g., optical depth, temperature
and cooling breaks will remain dependent on the actual mass of the
object.  Importantly,  the X-ray luminosities of both sources ($L_X/L_{\rm
  Edd}\equiv\ell_X\sim10^{-6}$) imply similar $\dot m$ \citep[see,
e.g.][]{Plotkinetal2012}, a necessary requirement for this
exploration.

\subsection {Fitting Methods} \label{sec:fit_method}

Given the complexities of both the data and the spectral model, we did
not expect to obtain straightforward fits with a reduced $\chi^2$
value of $\approx 1$ using simple Gaussian statistics.  We must
consider the fact that the error bars in BHBs represent
statistical errors on a near-simultaneous measurement, while for an
LLAGN we resolve ``waves'' of variability at levels of $\sim20\%$
typical for all bands \citep[e.g.,][]{Hoetal1999}. Such variability
would be averaged out over the much shorter BHB time scales (see the
discussion in \citealt{Markoffetal2008}). Direct comparison of errors
across broad energy bands and across mass scales therefore may be less
meaningful.  Nevertheless we do require some form of quantitative
measure of the quality of the spectral model descriptions, with a
means of judging the relative merits of different choices in model
assumptions and parameter values.  To this end, we have developed
exploratory methods to treat the data and perform the fits.

We are concerned with both the relative flux normalizations and
statistical weighting of individual observational bands.  As
differences can arise from cross-calibration uncertainties, we allow
for the usual fitted constant between spectra from different X-ray
satellites (see \citealt{Plucinskyetal2012}).  To account for delays
among energy bands and the lack of strict simultaneity among the
observations, as well as allow for systematic uncertainties between
instruments in different energy bands, we further adopt fractional, as
opposed to statistical, error bars for the non-X-ray data.  For V404,
we replace the non-X-ray statistical error bars with 5\% fractional
error bars.  (Larger error bars resulted in the few radio points
exerting too little statistical influence over the fits, smaller error
bars resulted in larger fit statistics regardless of fit parameters.)
For M81*, we replace the non-X-ray statistical error bars with 15\%
fractional error bars (i.e., comparable to the intrinsic radio
variability), except for the non-simultaneous IR/UV spectra where we
adopt 40\% fractional error bars.  For the UV data, there is some
debate whether these are detections of the emission from M81*, or are
merely upper limits to the central object emission
\citep[e.g.,][]{Maozetal2005}.  Adopting these large error bars thus
allows the HST and other non-simultaneous data to influence, but not dominate, the
model fits, and act as upper limits.  These choices admittedly contain
a degree of subjective judgement.  ``Best practices'' for combining
datasets from multiple, independent instruments remains an area of
active research, with some promising Bayesian methods allowing a more
formal approach to including priors for instrument systematics
\citep[see, e.g.,][]{Andersonetal2015}.  The focus of this work is to
first gauge whether tying parameters across such a large mass range in
these independent sources offers any viable solutions, with future
work devoted to refining parameter estimates of these models.

To fit the spectra, we begin with the usual approach of minimizing
$\chi^2$ with a fast algorithm, but we then extensively explore
parameter space via the use of an \texttt{ISIS} implementation
(described in detail in \citealt{MurphyNowak2014}) of the Markov Chain
Monte Carlo (MCMC) method of \citet{foreman:13a} and
\citet{goodman:10a}.  Parameter space is explored via 510 trial
``walkers'' which are evolved over a series of 3000 steps, only the
last 1000 of which are retained.  The resulting multi-dimensional
distribution of $5.4\times10^5$ parameter values are used to create
one- and two-dimensional histograms that then yield parameter error
bars and confidence contours.  The parameter set for the lowest
$\chi^2$ value found anywhere in this process is taken as the best-fit
model.

We start with the best fit parameters for the two degenerate classes
of models (synchrotron vs. SSC-dominated) fit to M81* from
\cite{Markoffetal2008}.  We then explore joint fits to the M81* and
V404 spectra, where we tie values of various parameter subsets for
the two sources.  As the values $T_e$, $T_{\rm in}$ and $f_{sc}$ are the most
obviously affected by local physical conditions, these particular
parameters are never tied.  Instead, we explore joint
fits where different subsets of the direct mass-scaling parameters,
$r_0$, $z_{\rm acc}$, $r_{\rm in}$, are tied.  We further explore
tying additional physical parameters, namely $p$ and $k$, that fall
within small ranges in prior studies of individual sources across the
mass range.

\section{Results}\label{sec:results}

\begin{figure*}
\begin{center}
 \includegraphics[width=0.45\textwidth]{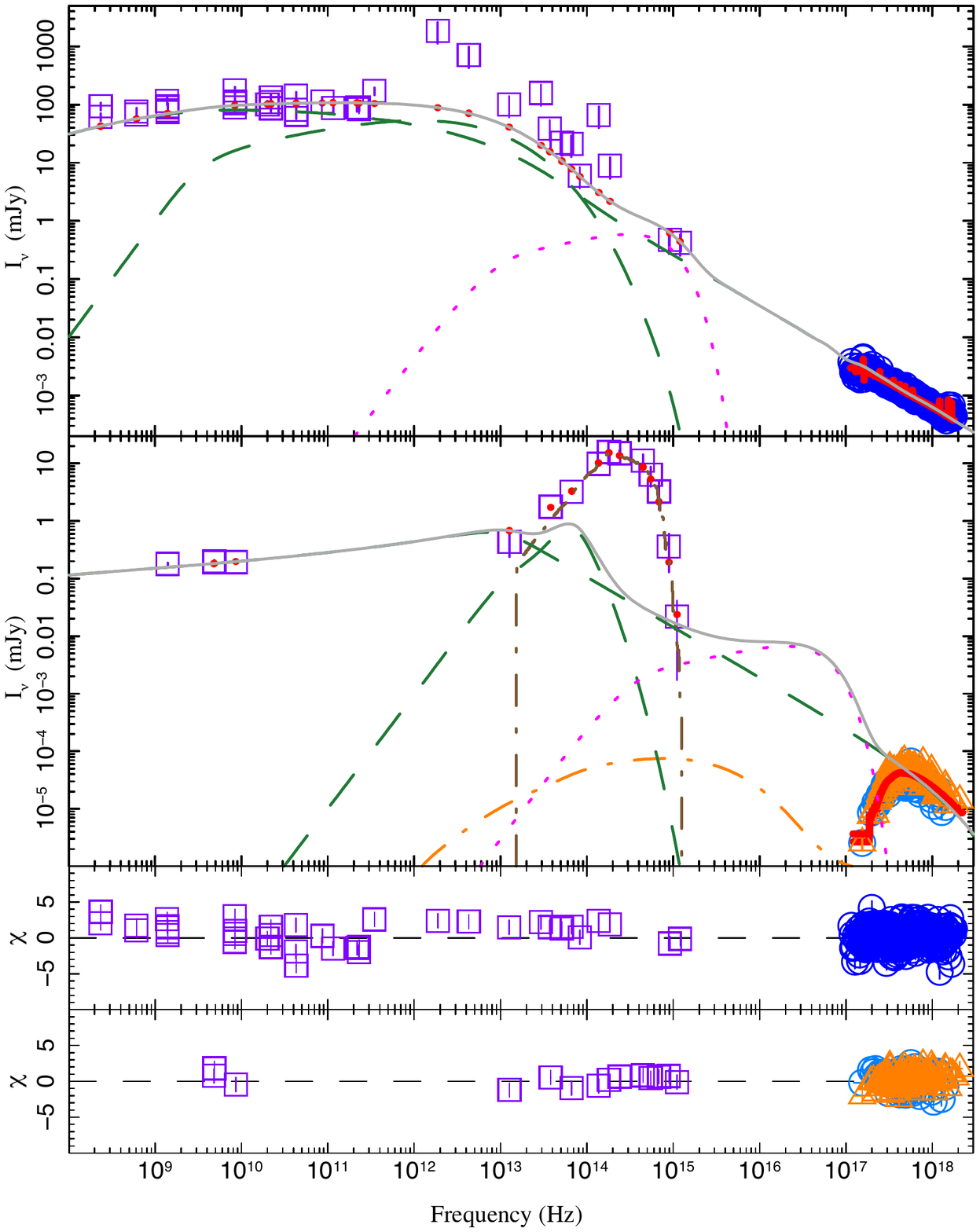}
 \includegraphics[width=0.45\textwidth]{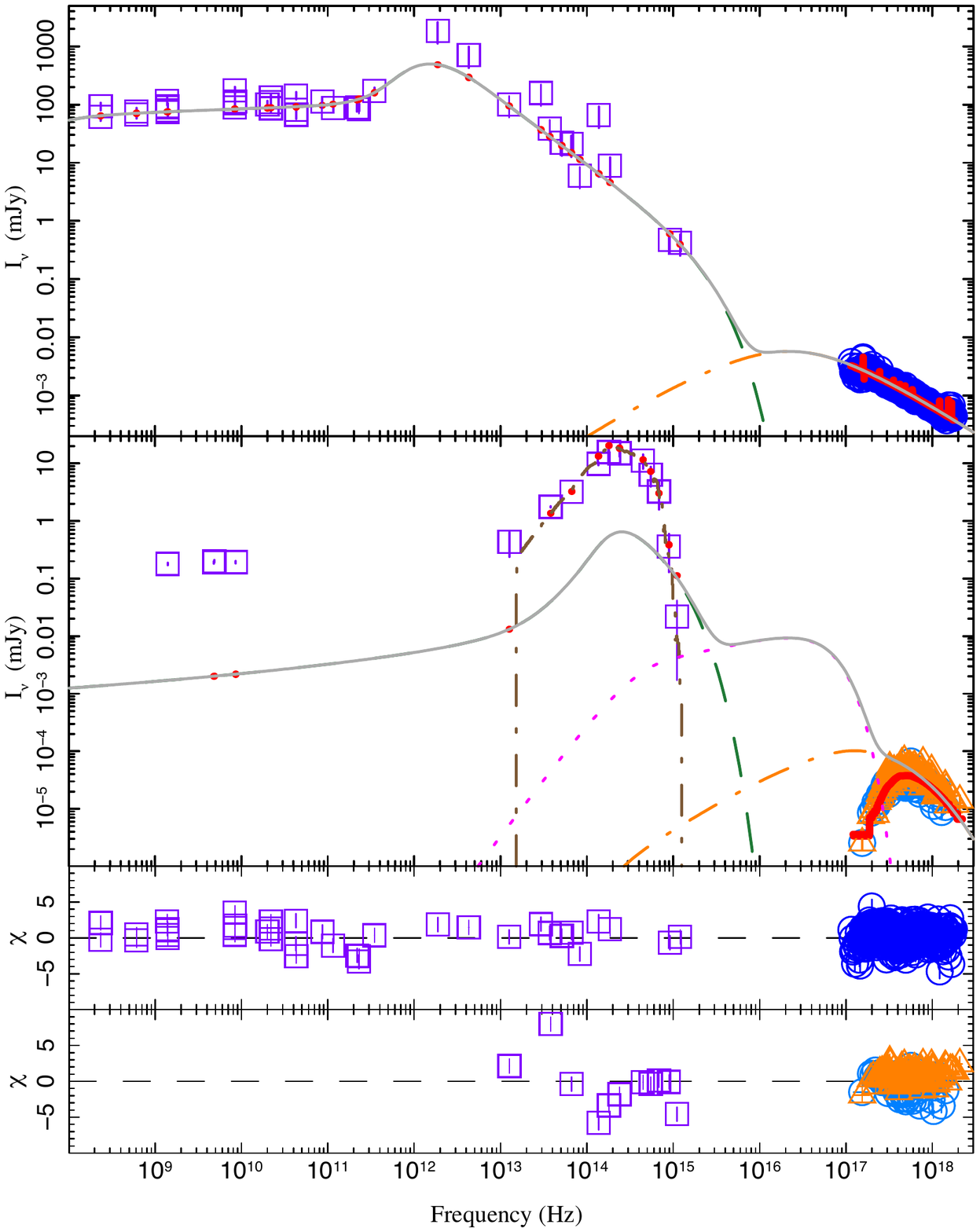}
\end{center}
\caption{Best fit synchrotron dominated model (left) and
  synchrotron self-Compton dominated model (right).  The top panels are the
  flux-corrected spectra for M81*, while the panels below are the V404
   flux-corrected spectra, and the bottom panels show the fit
  residuals.  Lines show the individual model components (light green/dashed:
  thermal synchrotron, dark green/dashed: non-thermal synchrotron, orange/dash-dot: 
  synchrotron self-Compton, magenta/dotted: multicolor blackbody disk, and grey/dash-dot: stellar component in the case of
  V404), all absent absorption.  The grey/solid line shows the total
  model, while the red/solid line and dots shows the model after
  forward folding through detector space, including absorption.
}
\label{fig:spectra}
\end{figure*}

\begin{center}
\begin{deluxetable*}{lllllllllllllc}
\tablecaption{Fit Parameters for Synchrotron Dominated and Compton Dominated Fits\label{table}}
\tablewidth{0pt}
\tablehead{
 \colhead{Source} & \colhead{$\rm N_{H}$} &  \colhead{$N_{\rm j}$} &  \colhead{ $T_{in}$} 
                  &  \colhead{$T_e$} & \colhead{$h_0$} & \colhead{$f_{sc}$} 
                  &  \colhead{$\gamma_{\rm max}$} &  \colhead{$p$} &  \colhead{$k$} 
                  &  \colhead{$r_{in}$} &  \colhead{$r_0$} &  \colhead{$z_{sh}$} 
                  & \colhead{$\chi_\nu^2/$DoF} \\
   &  \colhead {$\rm (10^{21}~cm^{-2})$} & \colhead {$(10^{-5})$} & \colhead{$(10^{4}~{\rm K})$}
   & \colhead{$(10^{11}~{\rm K})$} & \colhead{$(r_0)$} & \colhead{$(10^{-4})$} 
   & \colhead{$(10^2)$} & &
   & \multicolumn{3}{c}{$(GM/c^2)$}  
}
\startdata
${\rm M81^*}$ & $0.31^{+0.11}_{-0.03}$ &  $0.36^{+0.05}_{-0.04}$ & $1.6^{+0.3}_{-1.3}$ 
   & $16.6^{+1.6}_{-0.7}$ & 5\tablenotemark{a} & $47^{+8}_{-12}$ 
   & \nodata & \nodata & \nodata 
   & \nodata & \nodata  & \nodata 
   & \nodata \\
\noalign{\vskip1pt}
V404 & $8.8$\tablenotemark{a}  &  $17^{+20}_{-5}$ & $106^{+5}_{-37}$ 
   & $0.09^{+0.02}_{-0.02}$ & 1.5\tablenotemark{a} & $0.8^{+0.7}_{-0.2}$ 
   & \nodata & \nodata & \nodata 
   & \nodata & \nodata & \nodata 
   & \nodata \\
\noalign{\vskip1pt}
Joint & \nodata & \nodata & \nodata 
   & \nodata & \nodata & \nodata 
   & \nodata & $2.74^{+0.05}_{-0.01}$ & $0.73^{+0.53}_{-0.07}$
   & $1.1^{+8.7}_{-0.1}$ & $65^{+10}_{-3}$ & $305^{+15}_{-21}$ 
   & 1257/582 \\
\noalign{\vskip1pt}
\noalign{\hrule}
\noalign{\vskip1pt}
\noalign{\vskip1pt}
${\rm M81^*}$ & $0.03^{+0.02}_{-0.01}$  & $0.30^{+0.06}_{-0.04}$ & ${0.07}^{+0.49}_{-0.02}$ 
   & $2.1^{+0.3}_{-0.3}$ & \nodata & \nodata 
   & $15^{+20}_{-7}$ & $3.22^{+0.10}_{-0.13}$ & $0.45^{+0.48}_{-0.15}$ 
   & \nodata & \nodata & \nodata 
   & \nodata \\
\noalign{\vskip1pt}
V404 & $8.8$\tablenotemark{a} & $0.65^{+0.31}_{-0.03}$ & $112^{+8}_{-11}$ 
   & $0.35^{+0.09}_{-0.05}$ & \nodata & \nodata 
   & $0.22^{+0.09}_{-0.08}$ & $2.47^{+1.28}_{-0.36}$ & $0.32^{+0.28}_{-0.02}$ 
   & \nodata & \nodata & \nodata 
   & \nodata \\
\noalign{\vskip1pt}
Joint & \nodata & \nodata & \nodata 
   & \nodata & $10.1^{+5.3}_{-0.1}$ &\nodata 
   & \nodata & \nodata & \nodata 
   & $1.1^{+3.1}_{-0.1}$ & $3.7^{+1.5}_{-0.1}$ & \nodata 
   & 2671/582
\enddata

\tablecomments{Fit parameters for the synchrotron dominated (top) and
  SSC-dominated (bottom) fits.  The model components are: blackbody
  emission from the accretion disk and/or star (magenta/dotted),
  thermal synchrotron (light green/dashed), post-accelerated
  non-thermal synchrotron (dark green/solid), inverse Compton/SSC
  (orange/dash-dot).  Note in the SSC-dominated fit, accelerated
  particles were injected at the base, thus $z_{\rm acc}$ and
  $f_{\rm sc}$ are not used, while $\gamma_{\rm max}$ is.  These
  parameters gave the lowest $\chi^2$ values for all parameter space
  explored, while error bars are the bounds that encompass 90\% of the
  one dimensional parameter histograms obtained from MCMC exploration
  of the model fit (see text).  Other fixed physical parameters: mass
  (M81: $7\times10^7~M_\odot$, V404: 12 $M_\odot$), distance (M81: 3.6
  Mpc, V404: 2.4 kpc), inclination (M81: $20^\circ$, V404:
  $56^\circ$), see \citep{Markoffetal2008,Hynesetal2009}.}
\tablenotetext{a}{Frozen parameter.}
\end{deluxetable*}
\end{center}

\begin{figure*}
\begin{center}
 \includegraphics[width=0.33\textwidth]{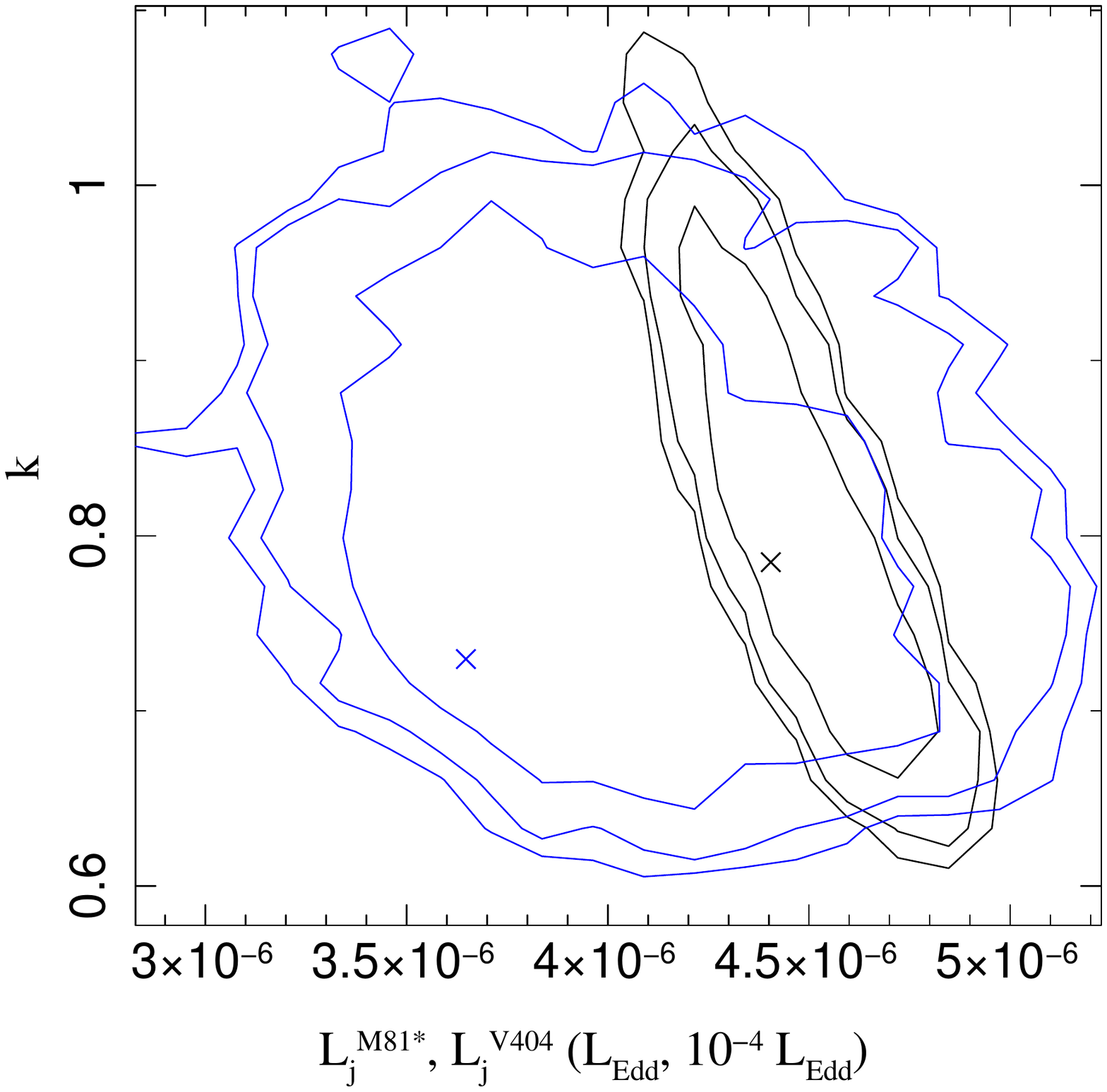}
 \includegraphics[width=0.33\textwidth]{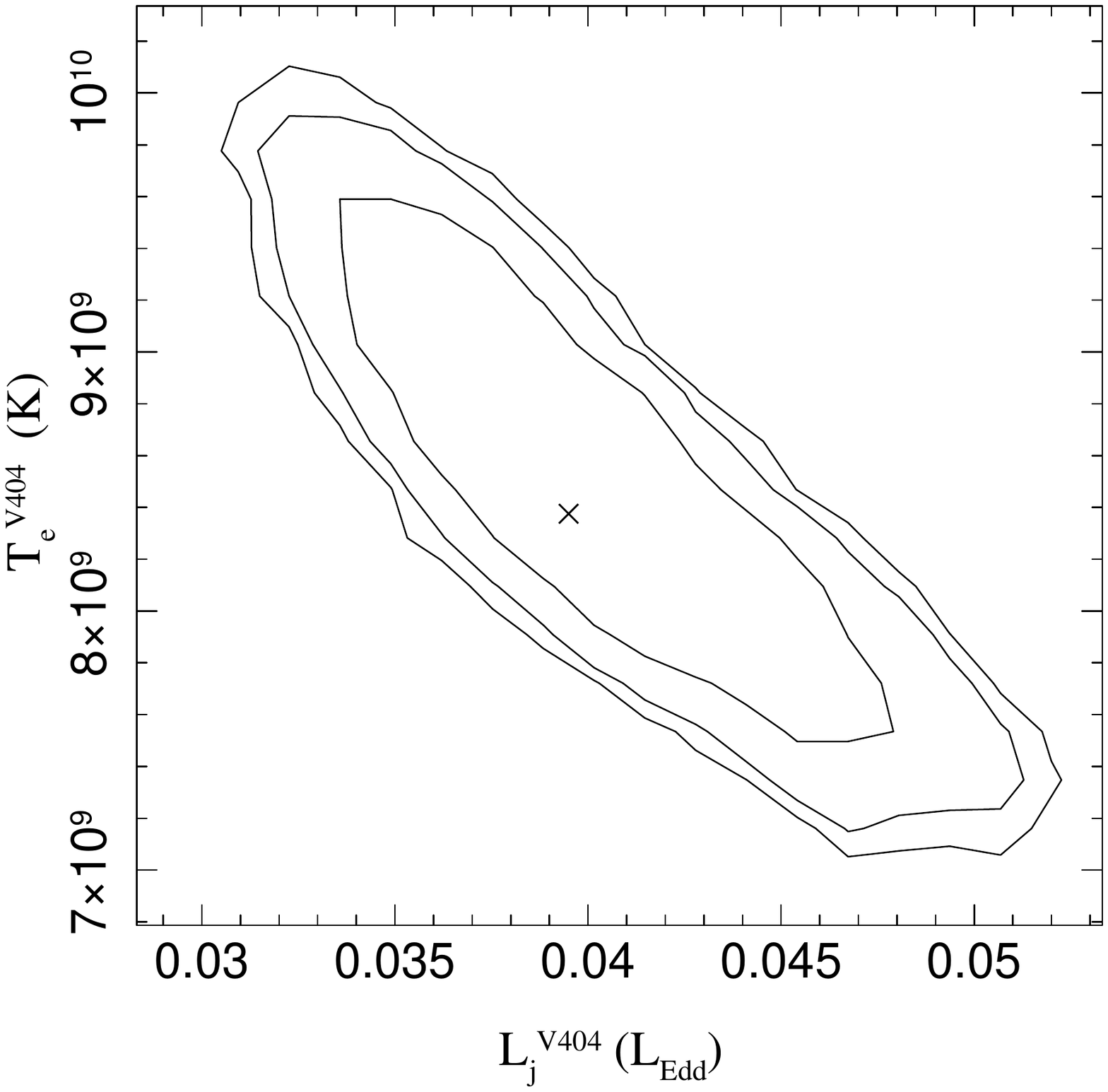}
 \includegraphics[width=0.33\textwidth]{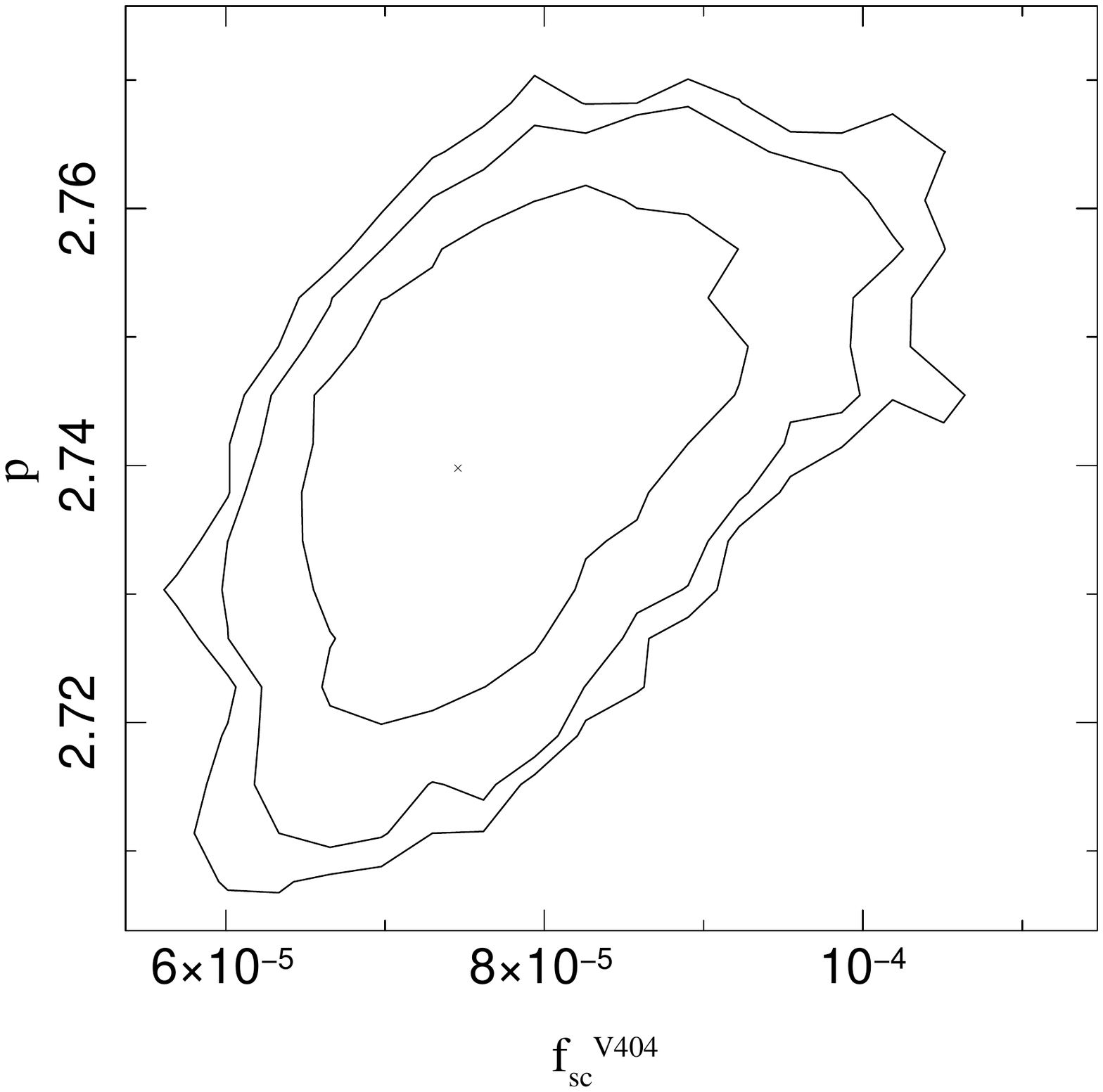}
\end{center}
\caption{The significant two-parameter correlations found
  via our MCMC exploration of parameter space for the
  synchrotron-dominated fits. Left: equipartition parameter $k$ (tied
  for both sources) vs. scaled injected jet power, $N_j$ (renormalized
  by $10^{-4}$ for V404).  Middle: Coronal temperature, $T_e$
  vs. scaled injected jet power for V404.  Right: powerlaw slope,
  $p$ of accelerated particle distribution (tied for both sources)
  vs. the plasma particle acceleration timescale parameter, $f_{sc}$,
  for V404.}
\label{fig:corrs}
\end{figure*}

In Table~\ref{table} we list the model parameters for the best fits
shown in Fig.~\ref{fig:spectra}, distinguishing between those free to
vary for each source and those tied together for a joint fit to both SEDs.

The synchrotron-dominated scenario is clearly the most successful,
providing a surprisingly good fit to both sources with almost half the
parameters tied -- including all relevant physical scales.  In
contrast, no SSC-dominated scenario could fit both sources in a
scaleable way.  While this result does not rule out SSC-dominated
scenarios, the idea that these two sources fall on the FP at similar Eddington fractions but via completely different
emission mechanisms seems less likely.  Even when decoupling some of
the tied parameters, we failed to find substantially improved fits.
Given that the synchrotron scenario not only had the best $\chi^2$,
but also allowed for the greatest number of tied parameters, we favor
the interpretation that synchrotron emission drives the FP correlation for at least the range
$\ell_X \sim 10^{-7}-10^{-6}$.

Compared to the best individual fits to M81* \citep{Markoffetal2008},
several parameters do not coincide within the errors to those found
here.  Specifically the joint fitting technique selects a slightly
hotter plasma injected within a larger jet base, and a slightly
steeper injected power law.  There are several potential reasons for
this difference, including the possibility that the earlier fits were
a local rather than global minimum since they were not obtained with a
MCMC approach.  It is worth noting that the M81*/V404 observations are
close to, but not exactly, at the same $\dot m$.  The individual data sets are
also not fully simultaneous.   Ultimately one would prefer to repeat this experiment
with fully simultaneous data sets at exactly the same $\ell_X$.  On
the other hand, the best fit parameter values still fall well within
the ranges found from earlier modeling of many individual sources.  Thus this new joint fitting approach does not fundamentally
change our ideas about the source physics or geometry, but rather
serves as a promising method to break the degeneracy between emission
scenarios.

The advantage of the MCMC approach is that with the multi-dimensional
probability distribution we can \textsl{a posteriori} explore all 120
possible two-parameter correlations.  This allows a new
level of insight into physical drivers of the FP as
well as pinpointing model degeneracy that needs to be addressed in
future work.  We find that the parameters for the synchrotron model
have well-determined means and errors as derived from their
one-dimensional histograms.  When examining two-dimensional
histograms, only a few parameters showed any degree of correlation
(see Fig.~\ref{fig:corrs}).  Several of these (not shown) are commonly
seen from fits to similar sources, e.g., correlations between fitted
neutral column and parameters affecting spectral slope.  Likewise for
M81*, there is a correlation between disk radius and temperature,
indicating that although a soft excess is required by the data, its
detailed properties are not well-determined.  Fig.~\ref{fig:corrs}
shows the 68\%, 90\%, and 99\% confidence contours from all
two-dimensional histograms where we see interesting correlations,
indicating either a physical relation or model degeneracy between
these parameters.  Both sources show a correlation (stronger for V404)
between the normalization power $N_j$ and the equipartition parameter
$k$.  This correlation indicates degeneracy in how the injected power
is divided between the radiating particles and the magnetic field.  As
$k$ is increased, putting more energy into the magnetic fields
respectively, less electrons are required for the same spectral fit,
resulting in somewhat lower power.  Less electrons can provide the
same energy density with a higher temperature, thus giving the
correlation seen in the middle panel.  Taken together these two
figures indicate a degeneracy between $N_j$, $k$ and $T_e$ in the
model, due to the parameterization of energy partition at the base of
the jets.  The rightmost panel shows a similar degeneracy between the
particle power-law index and$f_{\rm sc}$ on which the power-law cutoff
depends.  A harder value of $p$ can compensate for a lower cut-off up
to a point.

\pagebreak

\section{Discussion and Conclusions}\label{sec:conclude}

Our results support an emerging paradigm that the weakly accreting BHs
populating the Fundamental Plane can be treated as self-similar
objects, whose physical behavior is determined by accretion properties
rather than mass.  \textit{Specifically, we show that two BHs,
  separated by 7 orders of magnitude in mass but with comparable
  $\ell_X$, can be statistically described as ``self-similar'' in
  physical scale (in units of $r_g$)}.  For the more successful
synchrotron-dominated model, two additional parameters can also be
tied: the power-law distribution $p$, often thought to be universal
for a given acceleration process, and $k$, the partition of energy
density between magnetic fields and radiating particles.  The fact
that $k$ is roughly consistent with unity suggests that this parameter
could be eliminated with the assumption of equipartition.  The best
value for $p$ could imply either weak acceleration efficiency or very
efficient accelerations (such as from reconnection; e.g.,
\citealt{SironiSpitkovsky2014}) in a cooling-dominated regime.  The
SSC-dominated scenario does not achieve a good description of the
data, even with several additional parameters allowed to vary.
Interestingly, independent works suggest an interplay exists between
synchrotron and SSC as a function of $\dot m$, consistent with our
results.  E.g., \cite{Russelletal2010} empirically show that
synchrotron emission dominates the X-ray band around
$L_{\rm bol} \sim 10^{-4} - 10^{-3}$, while fits to LLAGN below
$L_X \sim 10^{-7} L_{\rm Edd}$ seem to prefer SSC radiation
\citep{Markoffetal2001,Plotkinetal2015,Prietoetal2015}.  The FP slope
does not seem to change despite this apparent transition
\citep{Corbeletal2013,Galloetal2014}, although the spectral index does
show softening below $\ell_X\sim10^{-5}$ \citep{Plotkinetal2013}.

The results of our study suggest that it is possible to exploit mass
scaling to break the longstanding degeneracies between the model
classes that persist for AGN \citep[see,
e.g.][]{HarrisKrawczynski2006} as well as BHBs
\citep[e.g.][]{Nowaketal2011}.  Compared to individual fitting, the
correlations found between parameters pinpoints the interplay between
parameter values due to model degeneracies as well as probing
meaningful physical relationships and the partition of energy between
magnetic, thermal and kinetic.  This new method thus opens the door to
several useful applications, such as using BHBs to infer conditions in
obscured regions deep in the hearts of galactic nuclei, or to study
processes that affect galaxy evolution over cosmological timescales.

Using mass-scaling for simultaneous joint/multiple fitting also has
the potential to constrain the SEDs of black holes with only sparse
data coverage, as well as better pegging the contribution of weak
accretion activity particularly in the mm/submm band of nearby
galaxies.  For instance, the discrepancy between the model and data in
the submm/OIR regime in Fig.~\ref{fig:spectra} is expected due to
galactic stellar and dust contributions \citep[e.g.][]{Bendoetal2010}.
We therefore plan to apply this new method to a larger sample of LLAGN with sub-arcsecond
aperture constraints on the galactic component
\citep[e.g.][]{Masonetal2012,Fernandez-Ontiverosetal2013} in future work.

Finally, a deeper understanding of why mass-scaling holds will
elucidate the respective roles of outer boundary conditions versus intrinsic
accretion flow physics, guiding the way towards more reliable
prescriptions of black hole feedback.

\acknowledgements SM is grateful to the University of Texas in Austin for its support, through a Tinsley Centennial Visiting Professorship.


\end{document}